\documentclass[fleqn,usenatbib]{mnras}

\usepackage{newtxtext,newtxmath}

\usepackage[T1]{fontenc}
\usepackage{ae,aecompl}

\usepackage{graphicx}	
\usepackage{amsmath}	
\usepackage{amssymb}	
\usepackage{mn2e-breakabs}

\def\etal{{\frenchspacing\it et al.}}

\title[First Release of AST3-1 Catalogue]{The First Release of the AST3-1 Point Source 
Catalogue from Dome A, Antarctica}

\author[Ma, Shang, Hu \etal]{
\parbox{\textwidth}{
Bin Ma,$^{1,2,3}$ 
Zhaohui Shang,$^{4,1,2}$\thanks{E-mail: \url{zshang@gmail.com}}
Yi Hu,$^{1,2}$ Keliang Hu,$^{1}$ Qiang Liu,$^{1}$
Michael~C.~B.~Ashley,$^{5}$
Xiangqun~Cui,$^{6,2}$ Fujia~Du,$^{6,2}$ Dongwei~Fan,$^{1}$
Longlong~Feng,$^{7}$ Fang~Huang,$^{8}$ 
Bozhong~Gu,$^{6,2}$ Boliang~He,$^{1}$ Tuo~Ji,$^{9}$ Xiaoyan~Li,$^{6,2}$
Zhengyang~Li,$^{6,2}$ Huigen~Liu,$^{10}$ Qiguo~Tian,$^{9}$ Charling~Tao,$^{11,12}$
Daxing~Wang,$^{6}$ Lifan~Wang,$^{7,13,2}$ Songhu~Wang,$^{14}$\thanks{51 Pegasi b Fellow}
Xiaofeng~Wang,$^{12}$ Peng~Wei,$^{10}$ Jianghua~Wu,$^{8}$
Lingzhe~Xu,$^{6}$ Shihai~Yang,$^{6,2}$
Ming~Yang,$^{10}$ Yi~Yang,$^{8,13,15}$ 
Ce~Yu,$^{16}$ Xiangyan~Yuan,$^{6,2}$ Hongyan~Zhou,$^{9}$
Hui~Zhang,$^{10}$ Xueguang~Zhang,$^{7}$ Yi~Zhang,$^{6}$ Cheng~Zhao,$^{12,1}$ 
Jilin~Zhou,$^{10}$
Zong-Hong~Zhu$^{8}$}
\vspace*{30pt}\\
$^{1}$National Astronomical Observatories, Chinese Academy of Sciences, 
Beijing 100012, China\\
$^{2}$Chinese Center for Antarctic Astronomy, Nanjing 210008, China\\
$^{3}$School of Astronomy and Space Science, University of Chinese Academy of Sciences, Beijing 100049, China\\
$^{4}$Tianjin Astrophysics Center, Tianjin Normal University,
Tianjin 300387, China\\
$^{5}$School of Physics, University of New South Wales, NSW 2052, Australia\\
$^{6}$Nanjing Institute of Astronomical Optics and Technology, Nanjing 210042, China\\
$^{7}$Purple Mountain Observatory, Nanjing 210008, China\\
$^{8}$Department of Astronomy, Beijing Normal University, Beijing 100875, China\\
$^{9}$Polar Research Institute of China, 451 Jinqiao Rd, Shanghai 200136, China\\
$^{10}$School of Astronomy and Space Science and Key Laboratory of Modern Astronomy and Astrophysics in Ministry of Education, \\
Nanjing University, Nanjing 210093, China\\
$^{11}$Aix Marseille Univ, CNRS/IN2P3, CPPM, Marseille, France\\
$^{12}$Physics Department and Tsinghua Center for Astrophysics (THCA), 
Tsinghua University, Beijing, 100084, China\\
$^{13}$George P. and Cynthia Woods Mitchell Institute for Fundamental
Physics \& Astronomy, Texas A. \& M. University, \\
Department of Physics and Astronomy, 4242 TAMU, College Station, TX 77843, USA\\
$^{14}$Department of Astronomy, Yale University, New Haven, CT 06511, USA\\
$^{15}$Department of Particle Physics and Astrophysics, Weizmann Institute of Science, Rehovot 76100, Israel\\
$^{16}$Tianjin University, Tianjin 300072, China
}

\date{Accepted XXX. Received YYY; in original form ZZZ}
\pubyear{2018}

\begin{document}
\label{firstpage}
\pagerange{\pageref{firstpage}--\pageref{lastpage}}
\maketitle

\begin{abstract}

The three Antarctic Survey Telescopes (AST3) aim to carry out time domain
imaging survey at Dome A, Antarctica. The first of the three telescopes (AST3-1) was
successfully deployed on January 2012.  AST3-1 is a 500\,mm aperture modified
Schmidt telescope with a 680\,mm diameter primary mirror. AST3-1 is equipped with a SDSS $i$ filter and a 10k $\times$ 10k
frame transfer CCD camera, reduced to 5k $\times$ 10k by electronic shuttering, resulting in a 4.3 deg$^2$ field-of-view.  To
verify the capability of AST3-1 for a variety of science goals, extensive commissioning
was carried out between March and May 2012.  The commissioning included a survey 
covering 2000 deg$^2$ as well as the entire Large and Small Magellanic Clouds. 
Frequent repeated images were made of the center of the Large Magellanic Cloud, a selected exoplanet transit field, and fields including some 
Wolf-Rayet stars.
Here we present the data reduction and 
photometric measurements of the point sources observed by AST3-1.  We have achieved 
a survey depth of 19.3\,mag in
60 s exposures with 5\,mmag precision in the light curves of bright stars.
The facility achieves sub-mmag photometric precision under stable survey conditions, 
approaching its photon noise limit.  These results
demonstrate that AST3-1 at Dome A is extraordinarily competitive in
time-domain astronomy, including both quick searches for faint transients and
the detection of tiny transit signals. 

\end{abstract}

\begin{keywords}
techniques: image processing -- methods: observational --
methods: data analysis
\end{keywords}

\section{Introduction} 

Small aperture telescopes with wide fields-of-view (FoV) have long played a
prominent role in time-domain astronomy.  Numerous projects have achieved
significant success in searching for rare transient events, such as near-Earth 
asteroids (NEAs), Potentially Hazardous Asteroids (PHAs), 
supernovae (SNe), gamma-ray bursts (GRBs) and tidal disruption events (TDEs), 
e.g., PTF \citep{2009PASP..121.1395L} and 
Pan-STARRS \citep{2002SPIE.4836..154K, 2016arXiv161205560C}.
The photometric monitoring of large sky areas also provides very valuable
data sets for variability studies of stars and active galactic nuclei (AGN). 

\citet{2016PASP..128h4501B} compares the survey capabilities of existing and 
planned projects, among which telescopes with a diameter less than one meter are 
also competitive. The Catalina Real-time Transient Survey (CRTS, 
\citealt{2009ApJ...696..870D}) has repeatedly scanned the sky for nearly a 
decade and detected thousands of SNe. Many areas of astronomy have benefited from such a 
dataset of long-term light curves, e.g., the finding of a possible 
supermassive black-hole binary in a quasar \citep{2015Natur.518...74G}.
The All Sky Automated Survey for SuperNovae (ASAS-SN, \citealt{Shappee2014}), 
deploys telescopes with 14-cm aperture lenses, and has discovered more than 500 
bright SNe since 2013, including the most luminous SN candidate ever found 
(\citealt{2016Sci...351..257D, 2017MNRAS.466.1428G}; however, 
\citealt{2016NatAs...1E...2L} and \citealt{2017ApJ...836...25M} suggest it is a TDE event).
Recently, there has been growing interest to develop transient surveys using 50-cm 
class telescopes. In Hawaii, the Asteroid 
Terrestrial-impact Last Alert System \citep[ATLAS,][]
{2011PASP..123...58T, 2018arXiv180200879T} 
consists of two 50-cm telescopes with 30 deg$^2$ FoVs. ATLAS automatically 
scans the entire accessible sky several times every night. A major science goal 
of ATLAS is to look for moving objects and provide warnings for killer asteroids. 
Since its first light in June 2015, ATLAS has 
discovered 127 NEAs, 17 PHAs, 9 comets, as well as more than one thousand 
SNe\footnote{\url{http://www.fallingstar.com/home.php}}. Meanwhile in La Silla, Chile, in the 
southern hemisphere, blackGEM \citep{2016SPIE.9906E..64B} is under construction. 
In Phase 1, it will be an array of three 65-cm telescopes, each with 2.7 
deg$^2$ FoV.  In Phase 2, blackGEM is proposed to be extended 
to 15 identical telescopes. Although ATLAS and blackGEM are equipped with the same CCD cameras, 
ATLAS has a pixel scale of 1\farcs86 to maximize its FoV, while blackGEM 
has a pixel scale of 0\farcs56 to obtain seeing-limited image quality and 
to push the survey depth down to $g \sim 22$.

Most ground-based transit surveys utilize small telescopes (aperture sizes
around 20\,cm) to search for exoplanets. Examples include WASP \citep{2006PASP..118.1407P}, HATNET
\citep{2004PASP..116..266B} and HATSouth \citep{2013PASP..125..154B}.  By 
surveying large areas of the sky to milli-mag (mmag) precision with rapid cadence, 
these surveys have discovered a number of exoplanets. Recently, the ongoing Next Generation 
Transit Survey \citep[NGTS,][]{2012SPIE.8444E..0EC, 2017arXiv171011100W} has employed
an array of twelve 20\,cm telescopes to find transiting Neptunes and super-Earths.

Near continuous monitoring of the sky enables comprehensive and rapid detections 
of transient phenomena that are intrinsically variable on short timescales. 
Studies of stellar variability also benefit from the
long-term continuous photometry that probes a wide range of frequencies.
For the highest-quality and most continuous datasets, a good site is essential.
The Antarctic Plateau has long been known as a premier astronomical
site since its atmosphere is extremely cold, dry, tenuous, and stable \citep{sau09}. Thus it is 
favorable for optical, infrared and THz observations. In particular, the
decreased high-altitude turbulence above the plateau results in reduced
scintillation noise \citep{ken06} thereby improving photometric and astrometric performance. The polar night in Antarctica provides the opportunity
for continuous observations of up to months, uninterrupted by the diurnal cycle
at temperate sites. Moreover, 
\citet{law04} reported a mean seeing of 0\farcs27 (median 0\farcs23) at Dome C above a low boundary layer \citep{2010PASP..122.1122B}, drawing the attention
of astronomers worldwide.  

In 2005, the 21st CHInese National Antarctic Research Expedition (CHINARE) arrived
at Dome A, the highest location of Antarctic plateau, for the first time.  Since then Chinese
astronomers have conducted site testing campaigns at Dome A in partnership with international
collaborators.  The results from various facilities have revealed that
Dome A has an atmospheric boundary layer as thin as 14\,m \citep{2010PASP..122.1122B}, a strong
temperature inversion above the snow surface but low wind speed 
\citep{2014PASP..126..868H}, low
water vapor \citep{2016NatAs...1E...1S}, low sky brightness and low cloud fraction
\citep{2010AJ....140..602Z, 2017AJ....154....6Y}.  In addition
to site testing, the first generation telescope, Chinese Small 
Telescope ARray (CSTAR, \citealt{2008SPIE.7012E..4GY, 2010RAA....10..279Z}),
continuously monitored an area of 20 deg$^2$ centered at the South Celestial Pole 
for three winters beginning in 2008.  The photometric precision from CSTAR reached 
4\,mmag after various efforts to correct the inhomogeneous effect of clouds 
\citep{2012PASP..124.1167W}, ghost images \citep{2013PASP..125.1015M} 
and diurnal effects \citep{2014RAA....14..345W}. CSTAR produced many studies 
of variable stars and exoplanets \citep[e.g.][]{2011AJ....142..155W, 
2015ApJS..217...28Y, 2015ApJS..218...20W, 2015AJ....149...84Z, 2016AJ....151..166O, 
2016AJ....152..168L, 2014ApJS..211...26W}.

Following CSTAR, the Antarctic Survey Telescopes (AST3) were conceived as the
second-generation optical telescopes at Dome A, and designed for
multi-band wide-field surveys, with each telescope operating with a different fixed filter.  
The AST3 telescopes would not only have larger apertures than CSTAR but would also have full 
pointing and tracking functions.  The first and the second AST3---AST3-1 \& 
AST3-2---were installed 
at Dome A in 2012 and 2015 respectively by the 28th and 31st CHINAREs. 
The third telescope, AST3-3, is 
under construction and will be equipped with a $K_{\hbox{\rm\scriptsize dark}}$-band near-IR camera 
\citep{2016PASA...33...47B,2016PASA...33....8L}.  

Here we present the first data release (DR1)
of photometric products from AST3-1 in 2012. These data have been used for a study of variable stars 
in one of the fields
\citep{2017AJ....153..104W}. We organize the paper as 
follows:  the telescope and CCD camera are introduced in \S2, the observations are described 
in \S3, the data reduction in \S4, we present 
the photometric results in \S5 and summarize the paper in \S6.

\section{AST3-1} 
The AST3-1 telescope, built by the Nanjing Institute of
Astronomical Optics \& Technology (NIAOT), has a modified Schmidt system
design \citep{2012MNRAS.424...23Y}. It has an entrance pupil diameter
of 500\,mm, a primary mirror diameter of 680\,mm, a focal ratio of $f/3.73$ and 
a large FoV with a diameter of $\sim 3\degr$.
The main features of AST3-1 include good image quality, a planar focal plane, 
reduced atmospheric dispersion, an absence of distortion and a compact structure. 
In the $r$- or $i$-bands, 80 per cent of the light energy of a point source is encircled 
within 1\arcsec.  The largest distortion across the FOV is 0.012 per cent,  
i.e. roughly 1\arcsec\ across the 3\degr\ diameter.  The tube is about half the length 
of a traditional Schmidt telescope. 
Unlike CSTAR, AST3-1 has full pointing and tracking components, as well as a
focusing system.  AST3-1 is powered by 
the PLATeau Observatory for Dome A\footnote{\url{ http://mcba11.phys.unsw.edu.au/~plato-a }}
(PLATO-A, \citealt{2010SPIE.7735E..40A}). As an evolution of the original 
PLATO, PLATO-A is a self-contained automated platform for supplying power of 
1\,kW continuously for a year, with Internet access provide by Iridium satellites.

AST3-1 is equipped with an SDSS $i$ filter and
the CCD camera is designed and manufactured by
Semiconductor Technology Associates, Inc.. The CCD has 10560 $\times$ 10560 pixels 
with a pixel size of
9\,\micron, corresponding to 1\arcsec\ in the focal plane of AST3. To avoid
the possible malfunction of a mechanical shutter, frame transfer mode was adopted to
terminate the exposure.  To do so,  the CCD is divided into two parts with
equal area: the frame store regions and the exposure area, giving an
effective FoV of $2.93\degr\times1.47\degr$, or $\sim$ 4.3 deg$^2$. There
are 16 individual readout amplifiers to accelerate the readout, which takes
40\,s in slow mode (100\,kHz), or 2.5\,s in fast mode (1.6\,MHz).
The CCD chip is cooled by a thermoelectric cooler (TEC) and takes advantage of
the low ambient air temperature at Dome A, which is about $-60^\circ$C on average
in winter\citep{2014PASP..126..868H}.  Detailed lab tests of the camera have been 
performed by \citealt{2012SPIE.8446E..6RM}, including linearity, the photon transfer curve, readout noise, 
dark current level and charge transfer efficiency. For example, the readout noise is 11\,$e^-$ 
in fast mode and is reduced to 4\,$e^-$ in slow mode.   
The camera was equipped with an engineering-grade CCD with
 many defective pixels and nearly half the area in 
one of the sixteen channels was damaged, resulting in an overall 3 per cent loss in 
effective FoV.

The hardware and software for the operation, control and data (COD) system were
developed by the National Astronomical Observatories, Chinese Academy of
Sciences (NAOC) \citep[][Shang et al. in preparation]{2012SPIE.8448E..26S, 
2016SPIE.9913E..0MH}.  The highly customized 
hardware consists of main control, disk array and pipeline computers. 
Each computer is duplicated to provide redundancy in order to minimize single points 
of failures. Attention to reliability was essential since the system has to run unattended during the observing season 
in austral winter. In order to ensure successful observations, we have overcome 
various technical difficulties encountered specifically in Antarctica.  
We have developed customized, stable, powerful but low power consumption 
computers and data storage arrays able to work in the harsh environment. 
Additionally, the software suite includes an automatic survey control, a 
scheduler \citep{2018RAA}, a data storage system, a real-time data processing 
pipeline and a robust photometry database system. The processing must be done on-site since the available Internet
bandwidth, and the data communication costs, prohibited bringing back more than a tiny fraction of the data in real-time.

\section{Observations in 2012 at Dome A} 
\begin{figure}
\includegraphics[width=8cm]{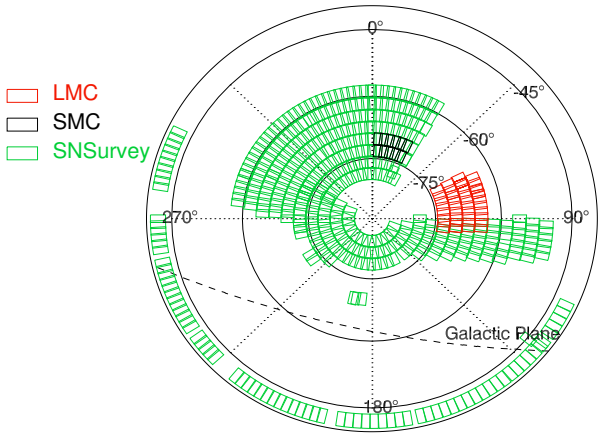}
\caption{The sky coverage of survey observations from AST3-1 in 2012.  
The small boxes denote the size and shape of the AST3 FoV while the colours 
indicate different surveys according to the legend. The dashed line is 
the Galactic Plane.
\label{fig:sky}}
\end{figure}

The first AST3 (AST3-1) was
deployed to Dome A, Antarctica by the 28th CHINARE team in January
2012\citep{2012SPIE.8444E..1OL} and became the largest optical telescope in
Antarctica. Installed during the bright sunshine of the polar day, AST3-1 had to wait for
operation until mid-March when twilight began. After 
tests of focusing, pointing and tracking, the telescope began commissioning
observations. Unfortunately the telescope stopped working on May 8 due to a
malfunction in the power supply system. During the March to May period, two major observing modes---the survey 
mode and the monitoring mode---were performed to verify the
performance of the telescope and CCD camera. In survey mode, the footprint covered 
roughly 2,000 deg$^2$ with 496 fields. The exposure time of each image 
was 60 s and in total more than 3,000 images were obtained.
The images were used as reference templates for searching for SNe and other transients; 
we refer to this as the SN Survey. The telescope also surveyed 50 fields covering the 
Large Magellanic Cloud (LMC) and 12 fields covering the
Small Magellanic Cloud (SMC). Fig.~\ref{fig:sky} shows the sky coverage of the survey mode. 
In a separate ``monitoring mode'', each of several fields were 
monitored for hours.  The monitored fields
include the center of LMC, a transit field where the exoplanets OGLE-TR-111b, 
113b and 132b are located, as well as the
Wolf-Rayet stars HD~117688, HD~136488, HD~143414 and HD~88500.
In addition to the surveys, AST3-1 observed 
many other targets such as stellar cluster, galaxies, quasars and 
follow-ups of SNe and GRBs.  Since these observations did not have many repeated exposures of the same fields,
we do not include these data here. 
Observation statistics for the survey and monitoring modes are shown in Table~\ref{tab:survey}
and Table~\ref{tab:monitoring}, respectively.

\begin{table*}
\begin{center}
\caption{Statistics of the AST3-1 survey mode observations in 2012}
\label{tab:survey}
\begin{tabular}{lrrrrr}
\hline
\hline
Object  &  No. of fields & Sky coverage  & No. of frames & Total exposure & No. of sources detected\\
 & & (deg$^2$) & & time (hr) & \\
\hline
LMC       &  50  &  200  &   664 & 10.60 &  3,038,210\\ 
SMC       &  12  &   50  &    55 &  0.9  &    227,608\\
SN Survey & 496  &  2000 &  3084 & 51.4  & 12,768,876\\
\hline
\end{tabular}
\end{center}
\end{table*}

\begin{table*}
\caption{Statistics of the AST3-1 monitoring mode observations in 2012}
\label{tab:monitoring}
\begin{center}
\begin{tabular}{lllrrr}
\hline
\hline
Object  &  RA & DEC  & No. of frames & Total exposure & No. of light curves\\
        & (deg) & (deg) & & time (hr) \\
\hline
LMC-center &  80.894  & $-$69.756  &  4183 & 58.6 & 815,589\\ 
Transit    & 163.9    & $-$61.5    &  3158 & 35.0 & 764,279\\
HD117688   & 203.3755 & $-$62.317  &   655 &  5.0 & 137,206\\
HD136488   & 231.0471 & $-$61.6771 &   660 &  4.2 & 148,798\\
HD143414   & 240.9557 & $-$62.6933 &  1442 &  7.6 &  66,284\\
HD88500    & 152.633  & $-$60.6451 &   591 &  2.2 &  41,578\\
\hline
\end{tabular}
\end{center}
\end{table*}

\begin{figure*}
\begin{center}
\begin{tabular}{l}
\includegraphics[width=8cm]{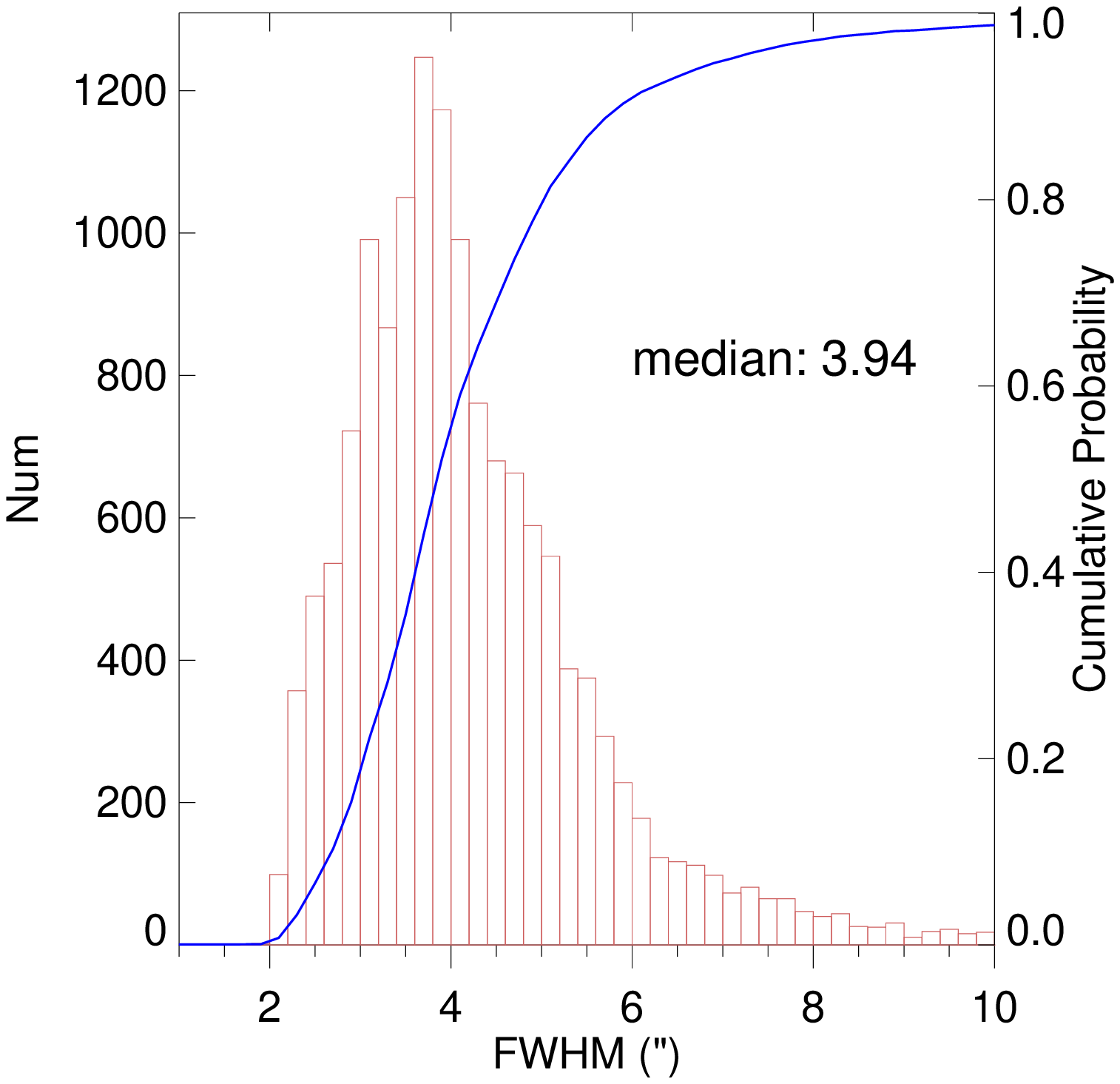}
\includegraphics[width=8cm]{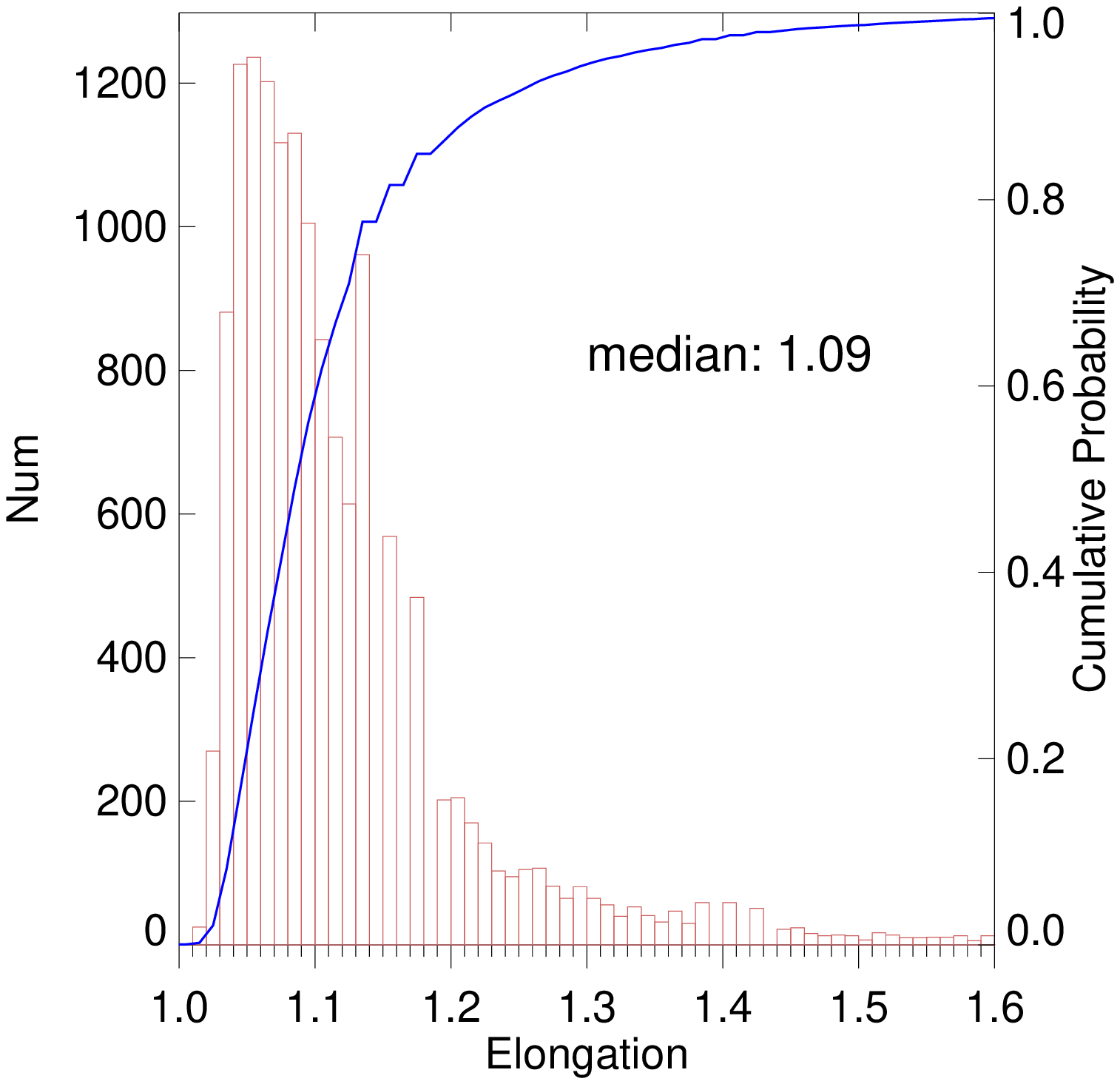}
\end{tabular}
\end{center}
\caption{Statistics for the image quality of the AST3-1 2012 commissioning observations. 
Left panel: FHWM and right panel: elongation.  
\label{fig:imqua}} \end{figure*}

AST3-1 was commissioned in unmanned operating mode at one of the most remote
sites on Earth, and with extreme site conditions.  Non-negligible instrumental effects needed to 
be modeled to process the data effectively.  The main issue was the irregular 
point spread function (PSF) resulting from significant tube seeing caused by the 
heat from the camera and the unstable tracking of AST3-1.  
As seen in Fig.~\ref{fig:imqua}, the full-width at half-maximum 
(FWHM) of the stellar images varied from 2\arcsec\ to 6\arcsec\ with a median of roughly 
4\arcsec.  The profiles of stars were elongated when unstable tracking occurred. 
The elongation, defined as the ratio of semi-major to semi-minor axis, had a 
median of 1.09 but exceeded 1.2 in some cases.  There was also a 
polar misalignment of about 0\fdg7 due to difficulties such as limited working time, manpower and resources at 
Dome A, while performing the installation 
process during daytime.   The pointing accuracy was improved to $\sim$ 2\arcmin\ after
TPoint correction\footnote{\url{http://www.tpointsw.uk/}} was applied to compensate 
for polar misalignment.

\section{Data Reduction} 
\begin{figure}
\includegraphics[width=8cm]{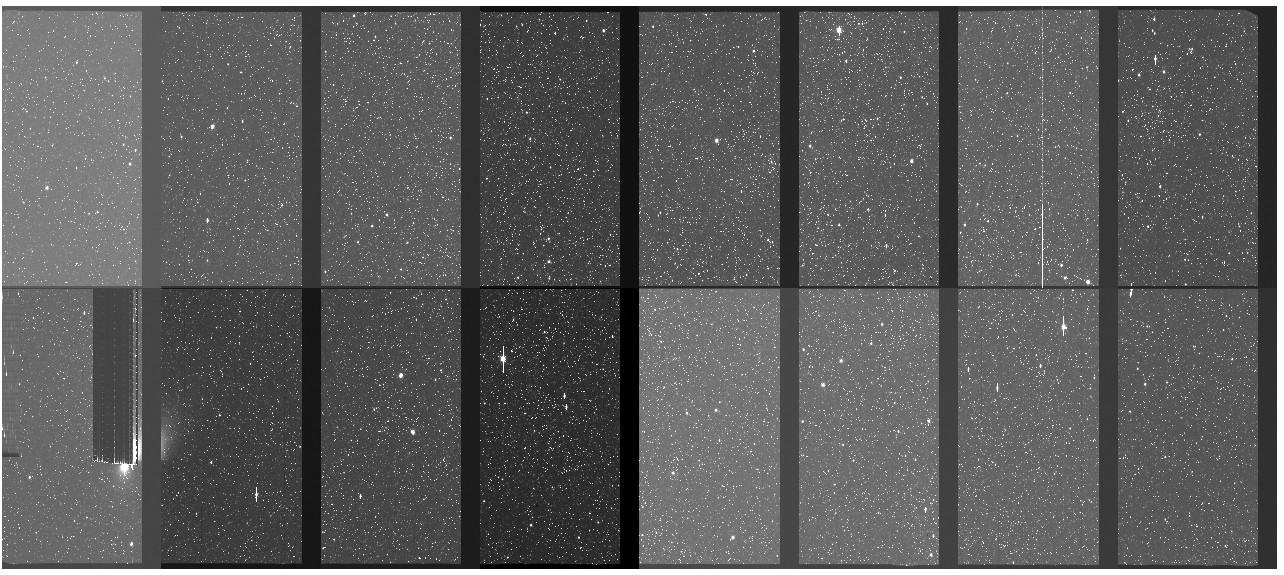}
\includegraphics[width=8cm]{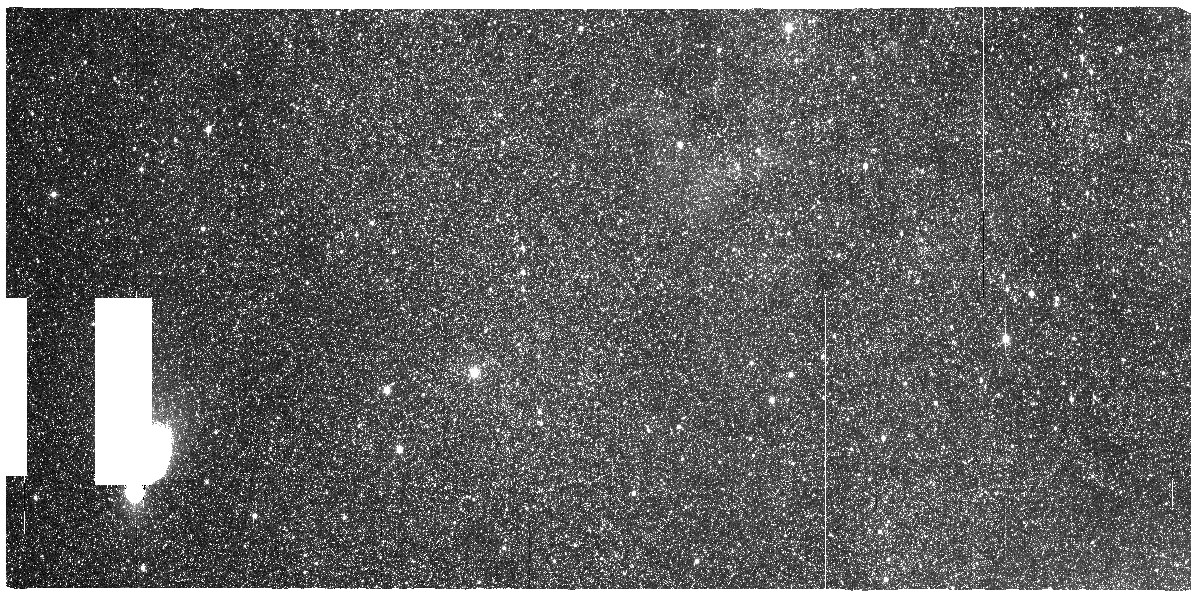}
\caption{Upper: an example of a raw frame from the Transit field where 16 readouts are 
obvious due to differing bias levels and over-scan. Lower: the preprocessed image from this example; 
two white regions in the left-lower readout mask the damaged part of CCD.
\label{fig:frame}}
\end{figure}

The raw data from AST3-1 were retrieved by the 29th CHINARE team in 2013.
The preliminary reduction of the raw science images involved corrections for 
CCD cross-talk, over-scan, dark current and flat-fielding. Fig.~\ref{fig:frame} 
illustrates an example of the reduction of a raw image in the Transit field. 
We then performed photometry 
on the preprocessed images and applied flux and astrometric calibrations on the extracted 
source catalogs.  Light curves were built by cross-matching 
catalogues in the same field.  Each step is detailed in the respective subsection 
below, particularly for custom methods used to derive the dark current and in
constructing the flat-field.

\subsection{CCD cross-talk} 
CCD cross-talk can occur when multiple outputs from the CCD are read out simultaneously.
If one amplifier reads a saturated
pixel, the pixels that are read simultaneously by the other amplifiers are
influenced.  This effect appears as ghosts in the image and can be either 
positive or negative depending on the CCDs.  Since there are sixteen readout amplifiers
in AST3-1's CCD camera, the cross-talk effect is significant.  Each saturated
star results in fifteen negative ghosts in other readout regions and the
stars that overlap these ghosts appear darkened.  To correct for this effect,
we compared the values between the ghosts and their surrounding pixels and
derived a uniform cross-talk coefficient $CT = 1.3 \times 10^{-3}$.  Then
for each readout, we search for every saturated pixel and add $65535 \times CT \sim 85$\,ADU
to the corresponding pixels at the same physical location in the other fifteen readouts. 

\subsection{Over-scan} 
Since AST3-1's camera has no mechanical shutter, even
a $0$~s exposure is exposed during the frame transfer period, which takes 434 ms, and consequently 
cannot be used as a bias frame.  Therefore, the median of over-scan columns on each 
readout was subtracted to remove the consequences of any voltage variations.

\subsection{Dark current} 
Dark currents in the AST3-1 camera were significantly high due to the heat dissipation 
from the CCD, which had a median temperature of $-46^\circ$C. While the dark current 
level is less than the sky brightness, the dark current non-uniformity
 can exceed the photon noise from the sky background.  
For example, the image a0331.116.fit taken at $t_{\hbox{\scriptsize CCD}} = 
-40^\circ$C has a median background of 620\,ADU.  The RMS of the 
background is 55\,ADU, which is roughly three times the photon 
shot noise (19\,ADU). Therefore it 
 is critical to correct the dark current. However, dark frames could not be
obtained during the observing season because of the shutterless camera and 
the unattended site in winter. We developed 
a simple but effective method to derive the dark current frame 
from the scientific images \citep{2014SPIE.9154E..1TM}. Here we briefly describe this
method.

The intensity $I$ of a pixel $(x, y)$ can be written as:

        \begin{equation}
        \label{eq:brightness}
I(x, y) = S + D(T) + \Delta d(T, x, y),
        \end{equation}
where $S$ is the sky background, $D(T)$ is the median dark current
level over the entire CCD at temperature $T$, and $\Delta d(T, x, y)$ is the deviation
from the median dark current at pixel $(x, y)$. The stars are ignored 
because they can be removed by a median algorithm when combining a large 
number of images from various fields. The sky brightness is adequately flat 
spatially after twilight \citep{2017AJ....154....6Y} and therefore is taken as a constant. 
In Eq.~\ref{eq:brightness} the sum of the first two
position-independent terms $S$ and $D(T)$
is practically the median value of the full image $I_0$.  We denote the later 
term as a fluctuation term, which describes the temperature and position-dependent 
effect of dark current. 
Considering two images taken with the same exposure time and temperature $T_0$, the sky 
brightnesses will in general be different. Recall Eq.~\ref{eq:brightness}, the two images have different
position-independent terms but the same position-dependent term:

        \begin{equation}
        \label{eq:dualbri}
        \begin{split}
I_1(x, y) &= I_{0,1} + \Delta d(T, x, y), \\
I_2(x, y) &= I_{0,2} + \Delta d(T, x, y).
        \end{split}
        \end{equation}
Supposing $I_{0,2}$ is brighter, $I_1(x, y)$ can be scaled to the equivalent
median level of $I_{0,2}$ by multiplying by the ratio $k \equiv
I_{0,2}/I_{0,1}$:

        \begin{equation}
        \label{eq:scale1}
I_1'(x, y) = k I_1(x, y) = I_{0,2} + k \Delta d(T, x, y).
        \end{equation}
Subtracting $I_2(x, y)$ from $I_1'(x, y)$ removes the constant
term and leaves only the fluctuation term. Consequently the 
non-uniformity of dark current at temperature $T$ can be calculated from the
image pair:

        \begin{equation}
        \label{eq:deltadark}
\Delta d(T, x, y) = \frac{I_1'(x, y) - I_2(x, y)}{k - 1} = \frac{k I_1(x, y) - I_2(x, y)}{k - 1}.
        \end{equation}
Adding the space-varying term $\Delta d(T,x,y)$ to the median level
$D(T)$, we can derive the dark frame
at temperature $T_0$ and exposure time $t_0$. Before applying the dark correction, the dark 
frame needs to be scaled to match the temperature and exposure time of the scientific 
images. However, it is insufficiently accurate to interpolate via the relationship between dark 
current level and temperature or time. Instead we use the dark current level 
of warm pixels as the scale parameter---these are high enough for precision 
determination and have an identical response to temperature and time as normal pixels.

This technique for correcting for dark-current non-uniformity has proved to be sufficiently robust.  
For the example above, the background RMS is reduced to 25\,ADU, 
which pushes the limiting magnitude 1\,mag deeper.

\subsection{Flat-fielding} 
 
We constructed a master flat-field frame by median-combining numerous
twilight frames.  However, the brightness of the twilight sky is not uniform, 
and the gradient varies with the Sun elevation and angular distance to the Sun.
Therefore, the varying gradients in individual twilight frames will introduce systematic
uncertainty in the median-combined frame.  We correct for the brightness gradient 
following the method discussed by \citet{2014SPIE.9149E..2HW}. 

AST3-1 obtained a total of 2,451 twilight frames during its 2012
observations.  We selected the images that had adequate sky intensity but were
still within the linearity range of the CCD, i.e., between 15,000 and 30,000\,ADU.  
We discarded images 
taken with CCD temperatures above $ -40^\circ$C, leaving
a total of 906 twilight images for further analysis.  
These frames were then median-combined to generate an initial master flat-field.  
Each twilight image was then divided by this master flat-field 
to remove gain variation and vignetting, yielding 
a relative gradient map to the initial master flat-field.  
We then fitted each gradient map with a two-dimensional
inclined plane $Z=a + bX +cY$.  By rejecting images which had
a large gradient or large fitting residuals, we obtained the final 
sample of 200 images.  Before recombining them, we removed the gradients 
by dividing then by their linear fitting plane.  This process was repeated for each 
readout separately rather than for the full frame, since a nonlinear gradient 
variation was observed across the entire frame.

Finally, we combined the gradient-corrected images to obtain the final 
master flat-field.  To compare the quality of the initial and final versions, we 
constructed the RMS image of two samples.  The initial RMS map 
showed a four-times increase from the center to the edge, while 
the final RMS map exhibits a rather uniform value across the entire frame at a 
level of 0.1 per cent.  
The accuracy of this flat-field is also confirmed by the 
photometric results in the next section.

\subsection{Photometry and Astrometry} 
Aperture photometry and astrometry was performed using
\textsc{sextractor}\footnote{\url{http://www.astromatic.net/software/sextractor}}
\citep{1996A&AS..117..393B} and
\textsc{scamp}\footnote{\url{http://www.astromatic.net/software/scamp}}
\citep{2006ASPC..351..112B}, respectively.  We set apertures
with radii of 2, 4, 6 and 8 pixels (or arcsecs at our plate scale of 1\arcsec\ per pixel). As a compromise between 
bright and faint stars, we adopted the magnitude with an aperture radius of 4\arcsec\ 
as the default magnitude. In addition, we derived a Kron-like
elliptical aperture magnitude MAG\_AUTO, which is usually a robust
estimator of total magnitude for galaxies.
The windowed position was adopted to calculate the centroid, which has been shown
to be as precise as a PSF-fitted position\citep{2007PASP..119.1462B}. 
To solve for the World 
Coordinate System, we adopt the Position and Proper Motions eXtended  
\citep[PPMX,][]{2008A&A...488..401R} as the reference catalogue for its small
size but high accuracy. The mean source density of PPMX is $\sim$ 440 deg$^{-2}$
and its typical 1-dimensional scatter is 40\,mas. Therefore, with nearly 2000
reference stars with precise coordinates in a typical AST3-1 FoV, we can ensure
accurate astrometric calibration. Our observed $1\sigma$ astrometric precision is
0\farcs1 in both RA and DEC, while the internal precision reaches 40 mas for
bright stars. This degree of precision with the astrometry is essential for image
registration when using difference image analysis (DIA) for transient object detection.

\subsection{Flux Calibration} 
We adopted the AAVSO Photometric All-Sky Survey (APASS
\footnote{\url{https://www.aavso.org/apass}}) DR9 catalogue  
\citep{2016yCat.2336....0H} as the reference to
calibrate AST3-1's magnitudes.  APASS is an all-sky survey conducted in $BVgri$
filters. The $i$-band limiting
magnitude reaches a depth of $\sim 16$. Compared to the secondary $ugriz$ standards 
\citep{2002AJ....123.2121S}, APASS $i$ magnitude has a RMS scatter of 0.039\,mag and no color 
trend \citep{2014AJ....148...81M}. For the survey fields, we
calibrate each frame to APASS. While for each monitored field, we only
calibrate the frame with the best image quality to APASS and then adopt this 
frame as a reference to calibrate the others. The best frame is selected 
as the one with the most detected sources, which results from a combination of effects such as FWHM, sky brightness, 
dark current level and extinction.

\begin{figure*}
\begin{center}
\begin{tabular}{c}
\includegraphics[width=8cm]{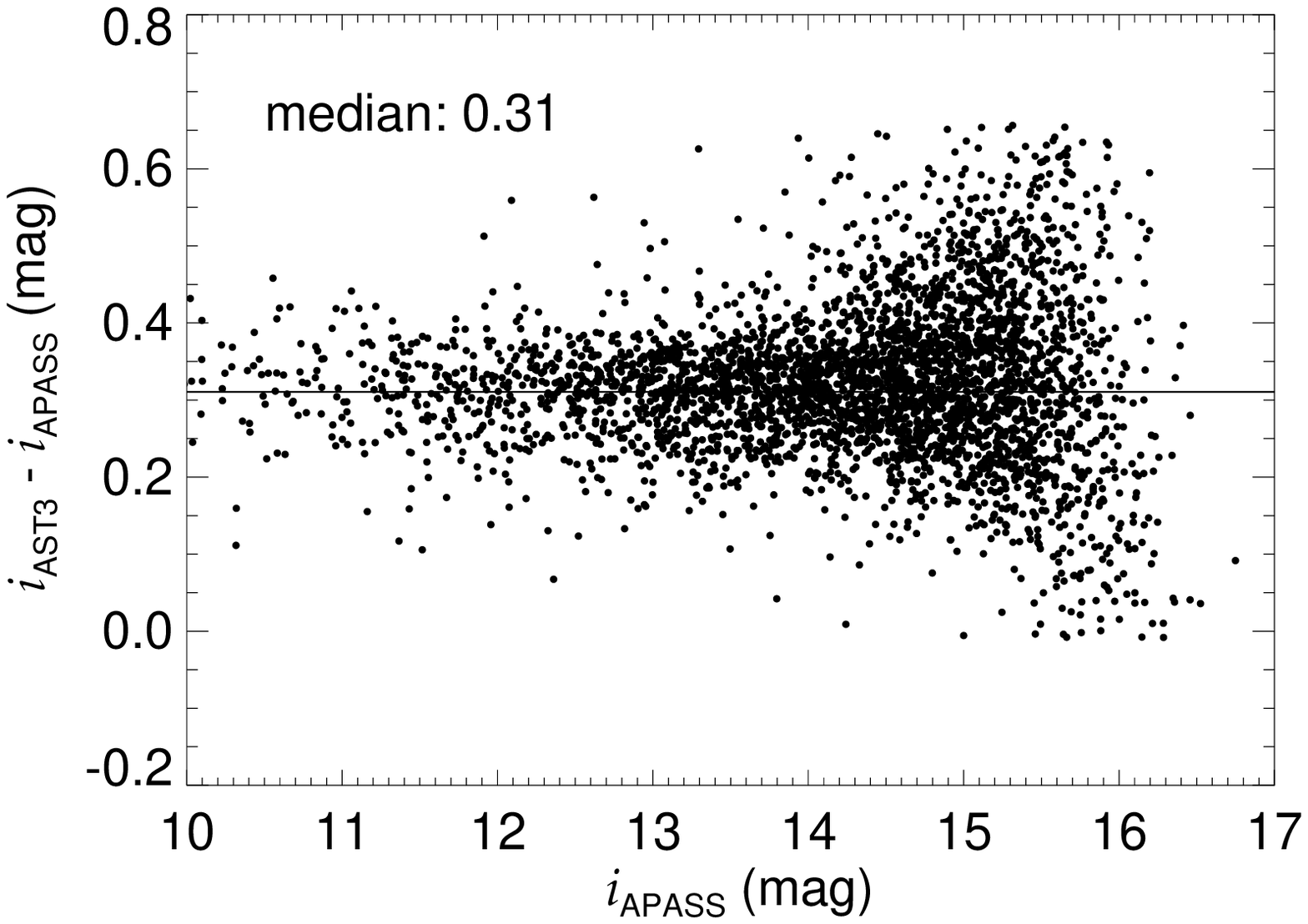}
\includegraphics[width=8cm]{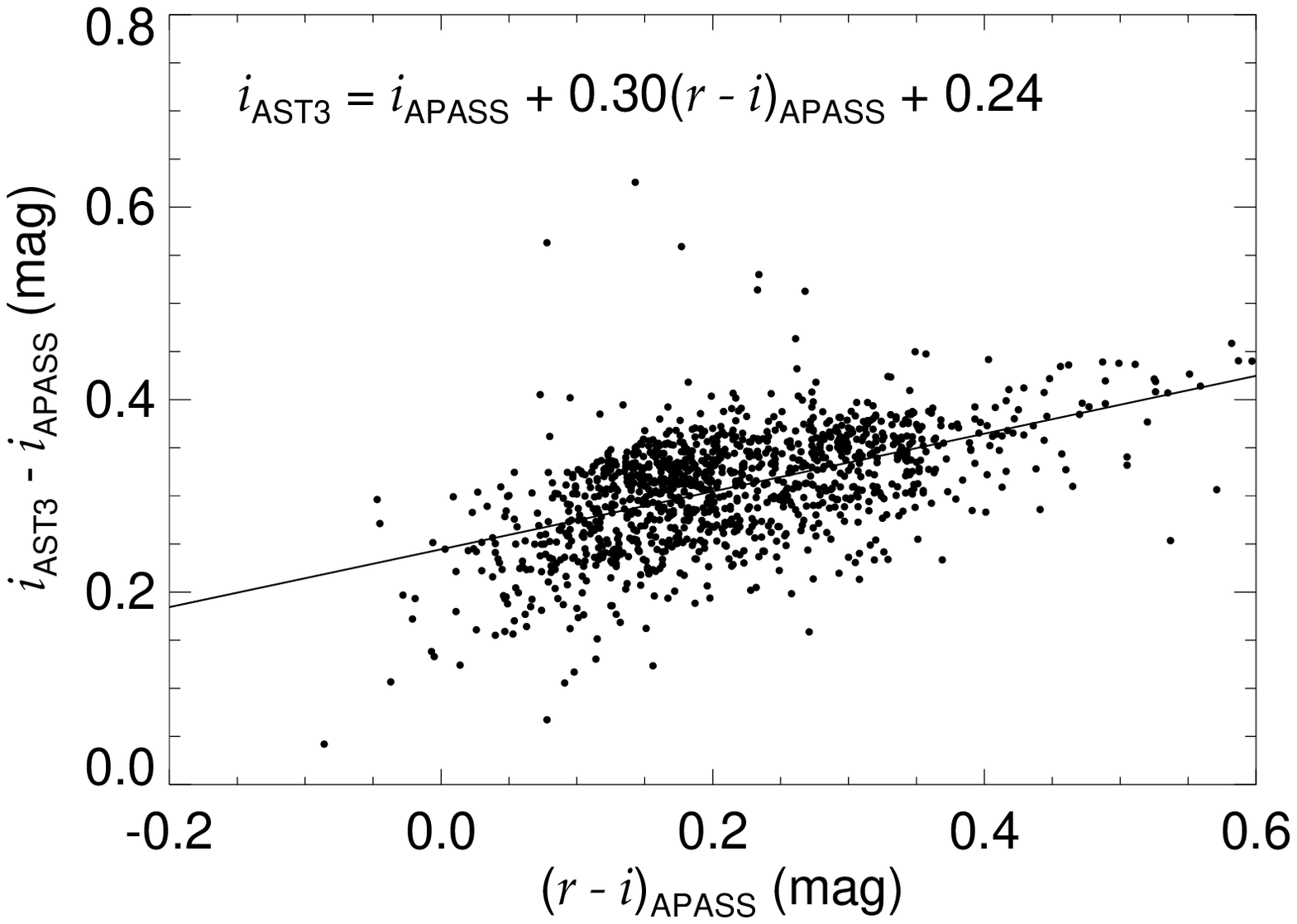} 
\end{tabular}
\end{center}
\caption{The difference between APASS $i$-magnitudes and AST3-1 instrumental 
magnitudes of 3,636 stars in an example frame a0420.433.fit as a function of (a) the 
APASS $i$-magnitudes shown in the left panel and (b) the APASS $r-i$ color for 
1,207 stars with $i_{\rm APASS}$ between 11 and 14 shown in the right panel.
The solid line in the left panel denotes the median value of the magnitude 
differences, which is adopted as the magnitude zero point.
And the solid line in the right panel is the best 
linear fit to the colour dependence.\label{fig:magzp}}
\end{figure*}

When absolutely calibrating an image, we calculate the photometric offsets from instrumental 
to available APASS $i$ magnitudes. The median offset of each frame 
was used as the zero point to calibrate the magnitude of all the sources in the frame. 
The left panel of Fig.~\ref{fig:magzp} presents the offset from an example image
(a0420.433.fit).  Stars fainter than 14th magnitude exhibit larger dispersions due 
to the limited depth of APASS.  Therefore, we only use 11--14\,mag stars to derive the
zero-point, which is 0.31\,mag with a standard deviation of
0.06\,mag in this example.  A colour-free zero-point is applied since there was only one
filter for AST3-1 in 2012. To investigate the effect of a colour term in the AST3-1 
photometric calibration, we also compared the magnitude differences as a function of 
associated APASS $r-i$ colour, shown in the right panel of 
Fig.~\ref{fig:magzp}(b).  The best linear fitting of the 
magnitude transformation is 
$$i_{\hbox{\scriptsize AST3}} = i_{\hbox{\scriptsize APASS}} + 0.3 (r-i)_{\hbox{\scriptsize APASS}} + 0.24 .$$
Many factors contribute to the difference between two systems, including 
telescope throughput, 
CCD quantum efficiency and atmospheric transparency. The low water vapour contents of the atmosphere above Dome A results in
significantly better transmission \citep{2012PASP..124...74S}, which can affect the effective filter curves. The colour coefficient 
is as large as 0.3. Therefore caution should be taken when comparing
AST3-1's magnitudes with other surveys, especially for extremely blue or red
targets. It is also critical when combining magnitudes from multiple facilities to
derive the light curve of a colour-varying source. 
For example, tens of telescopes worldwide contributed to the observations
of GW170817, the first optical counterpart of a gravitational wave event
\citep{gw170817}.  However, its $r-i$ colour varied from less than 0 to 
nearly 1\,mag over four days.  Consequently, large systematic errors of up to 0.3\,mag 
can be induced in the light curve 
if no colour correction is applied between the various telescopes that are nominally observing
in the same band.

\begin{figure} 
\includegraphics[width=8cm]{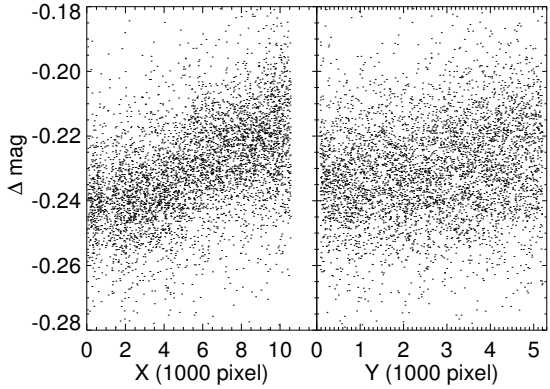} 
\caption{The obvious dependence of $\Delta$mag
between b0409.267.fit and a0410.316.fit on CCD 
(X,Y) positions for 4,956 stars with $S/N>$ 50.\label{fig:dmxy}} 
\end{figure}

In relative calibration for the same field, we find discernible position-dependent
variations in magnitude differences between two frames.  The reason is 
probably the variation of FWHM across the CCD and the application of an uniform 
size of aperture.
Fig.~\ref{fig:dmxy} illustrates an example of $\Delta$\,mag between 
a0410.316.fit and b0409.267.fit, which is the reference image in 
the field of HD~143414. The magnitude zero-point differs on the order
of 0.02~mag across the CCD.  To correct for this, we fitted a quadratic zero-point as 
a function of (X,Y) from using stars with $S/N> 50$, and adopt this to
calibrate all the stars.  This further reduces the observed scatter in the light curves for bright stars 
away from the CCD center compared to a constant zero-point.  However, we do not 
apply the same correction in absolute calibration due to the insufficient accuracy 
of APASS.

\section{Photometric accuracy}

\begin{figure} 
\includegraphics[width=8cm]{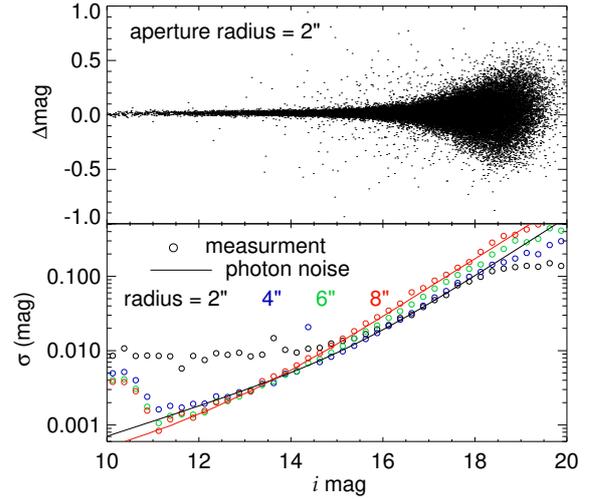} 
\caption{The upper panel presents the magnitude differences between two 
consecutive 60\,s exposures, a0330.104.fit and a0330.105.fit as a function 
of $i$-band magnitude, measured with a circular aperture of 2\arcsec\ in
radius. The lower panel shows the RMS calculated in each 0.2-mag interval and 
different colors represent the results obtained for different aperture radii. 
Solid lines indicate the expected error from photon noise for different 
cases. The performance of the small aperture at the faint end shows a
$5\sigma$ limiting magnitude of 18.7, while the precision from large apertures at
the bright end reaches 0.8\,mmag, roughly the photon noise limit. It implies 
that the systematic error is indiscernible.\label{fig:magerr}} 
\end{figure}

In order to estimate the photometric precision of single frame, we compared the magnitude
differences between two images and illustrate with an example in
Fig.~\ref{fig:magerr}.  The results are based on two consecutive 60 s images
a0330.104.fit and a0330.105.fit with FWHM $\sim$ 3\farcs7.  The upper
panel shows the differences of aperture magnitude using a 2\arcsec\ radius and
the lower panel shows the $1\sigma$ deviation of magnitude difference in
each 0.25\,mag interval. Over-plotted solid lines indicate the photon noise of 
the signal and the background calculated for various
apertures. The photon noise can be written as:
$$ \sigma =1.0857 \frac{\sqrt{(\pi r^2 S+F)/g}}{F} ,$$ 
where $r$ is the aperture
radius, $S$ is the sky background, $F$ is the source flux enclosed by the aperture $r$
and $g$ is the gain of CCD in electrons/ADU.  Here we do not consider the readout noise, which 
is negligible compared to the photon noise from the sky background.  Nor do we include 
scintillation noise since its level is about 0.1\,mmag \citep{ken06}, much is 
smaller than the photon noise of the brightest stars.  Furthermore, we compare 
the precisions using aperture radii of 2, 4, 6, and 8\arcsec.

As expected, smaller aperture provides more accurate measurements for the fainter range where
the noise is dominated by sky background, while larger aperture is more suitable
for bright stars where photon noise from the stars is dominant.  For all
apertures, the measured noise decreases when brighter and is almost
identical to the photon noise in moderate magnitude range.  However, it
exceeds photon noise for bright stars, tending to be a constant regardless of
decreasing photon noise, and starts to rise for stars brighter than $i
\sim$ 11, which are becoming saturated.  The larger the aperture is, the smaller
the deviation from photon noise.  The constant noise is caused by
systematic errors, including instability of the telescope and camera, variation of
the atmospheric transparency and seeing, as well as residual errors from the flat-field and
dark current corrections. In this example, the best $1\sigma$ precisions for four
apertures are 10, 1.6, 1.1 and 0.8 mmag,  respectively.  By comparing with
the photon noise it indicates that the systematic error is 10 mmag for
aperture of 2\arcsec, but is dramatically reduced to roughly 1 mmag for an aperture
of 4\arcsec\ and is even negligible for aperture of 8\arcsec.  Since flux within a radius
of 2\arcsec\ contributes to nearly half of the total flux, the systematic error would
be reduced by a factor of 2 at most if little noise comes from
annulus beyond 2\arcsec.  However, we observe a decrease by a factor greater than 10,
which is likely a result of the difficulty in apportioning fluxes to fractional pixels in the 2\arcsec\ aperture. 
In summary, the stability of the AST3-1
system is much better than 1 mmag during the two minutes when these two images were taken.

At the faint end when $i > 18.5$, the measured error tends to be flat due to the
lack of stars with $\Delta {\hbox{mag}} < 0$ that are detected in one frame, but
beyond the detection limit in the other frame.  This selection bias results in
the measured noise underestimating the real noise, so we take the photon
noise to define the photometric depth.  The $S/N = 5$ limiting magnitude is
18.7 with an aperture radius of 2\arcsec\ in this case.

These results demonstrate the capabilities of AST3-1 in both detection depth and
photometric precision.  A deeper detection limit would result from a sharper FWHM and
lower sky background.  For example, a FWHM of 2\farcs6 in another image pair,
a0331.116.fit and a0331.117.fit with similar sky brightness pushes the
$5\sigma$ limiting magnitude down to $i \sim 19.3$. Here we remind the reader 
that although the seeing at Antarctic sites at a few meters above the boundary
layer can be exceptional and reach values of 0\farcs3, the surface 
seeing is quite poor. For instance, the median seeing measured at the ground 
surface at Dome C is about 2\arcsec\ \citep{2009A&A...499..955A}. On the other 
hand, the sky background in these
examples is about 19.5\,mag arcsec$^{-2}$ and would be one mag darker in a
moonless polar night \citep{2010AJ....140..602Z}, which AST3-1 observations did not
sample as a result of the power supply problem in early May 2012.
Consequently, we infer that AST3 is able to approach a
depth of $i \sim 20$ in a 60\,s exposure during dark nights under
seeing-limited conditions within the turbulent boundary layer.  

On the contrary, a sharper FWHM would degrade the 
photometric precision for bright stars.  For example, the precision in an image
pair of a0331.116.fit and a0331.117.fit is degraded to 2\,mmag.  However,
FWHM larger than 4\arcsec\ hardly improves the accuracy, which is usually
dominated by systematic errors, but it extends the dynamic range to brighter 
stars.  Therefore we can optimize the observational strategies for the transit
survey mode according to the target magnitude ranges.

\begin{figure} 
\includegraphics[width=8cm]{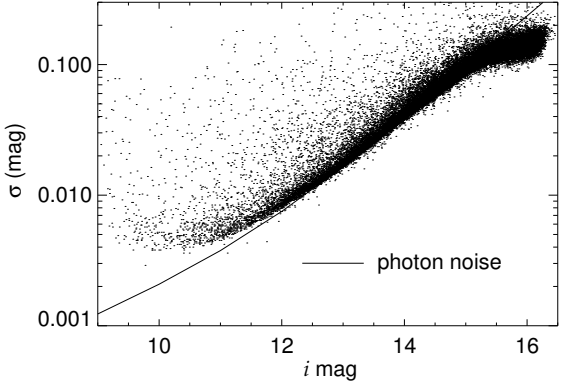} 
\caption{Typical dispersions for AST3-1 $i$-band light curves as a function 
of stellar brightness in field HD~143414.  The solid line shows ideal
photon noise under typical observing conditions. The
RMS uncertainties for bright stars are limited to $\sim 5$\,mmag due to 
systematic errors and relative calibration error.\label{fig:lcrms}} 
\end{figure}

For long-term monitoring rather than just two continuous images, the systematic
errors become predominant in light curves of bright stars, constraining
the detection of tiny variabilities such as exoplanet transit signals.
For instance, Fig.~\ref{fig:lcrms} shows a diagram of the RMS scatters in the 
light curves versus instrumental magnitude in the field centered on a Wolf-Rayet 
star HD~143414. This field was monitored on 8, 9, 10, 11 April and 5,
6 May for one to four hours on each night. We have obtained high-precision 
photometry with an RMS accuracy
of $\sim 5$\,mmag at the bright end. We also present the theoretical 
photon-noise limit in a typical observing condition (an integration
time of 10\,s, a FWHM of 4\arcsec, sky background of 1000\,ADU due to full Moon and
an extinction of 0.7\,mag due to cloud or frost on the mirror).  On account of
these conditions, photon noise of the brightest stars is 2\,mmag.  
The systematic errors arise
from  variations in exposure time, FWHM, sky brightness, transparency and
position on CCD, as well as errors induced in image corrections and
calibrations.  The photometric accuracy may be further improved with more careful
de-trending by those who are interested in these data.  Similar result 
has been also obtained for the Transit field 
\citep[see Fig.~1 in][]{2017AJ....153..104W},  while for crowded fields
such as the LMC-center, the photometric accuracy for bright stars is mainly
limited by the difficulty of using aperture photometry on blended sources.  To 
improve the precision, PSF photometry is desired and will be performed in the future. 
\\

\begin{figure} 
\includegraphics[width=8cm]{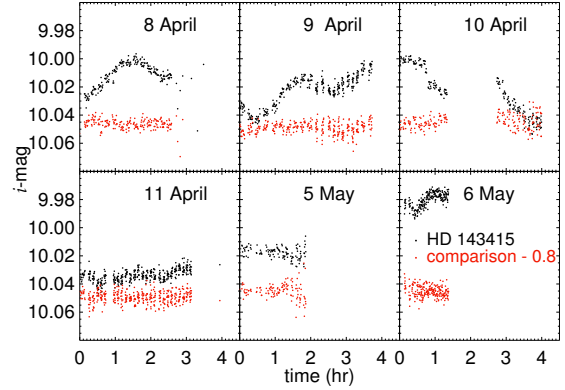} 
\caption{Light curves of 
HD~143414 (black dots) and a comparison star (red dots) on six days
Time zero denotes the first exposure of the star on each day.  The
magnitude RMS of the comparison star reaches 6\,mmag, and HD~143414 exhibits
obvious aperiodic variability on timescale of shorter than an
hour.\label{fig:lc}} 
\end{figure}

In Fig.~\ref{fig:lc}, we plot example light curves for HD~143414, one of
our Wolf-Rayet star targets.  The epoch of the
observations are labeled. A photometric precision of better than 6\,mmag is indicated 
by the scattering in the light curves of the comparison star, which is 210\arcsec\ away
and $\sim 0.8$\,mag fainter.  The light curve of HD~143414 exhibits
diverse variability on individual days and variabilities shorter than
an hour are obvious. Detailed studies of stellar variabilities in the Transit field 
from the AST3-1 data set can be found in \citet{2017AJ....153..104W}.

Finally, we compared the AST3-1 photometry with the Southern 
Extension of $ugriz$ standard 
stars\footnote{\url{http://www-star.fnal.gov/}} \citep{2007ASPC..364...91S}. 
This comparison enables a sanity check of the absolute photometric calibration 
of AST3-1 observations. The standards are located in
a grid of fields spaced roughly every two hours of RA along Dec of $-30\degr$ 
and $-60\degr$, as well as some special fields.  Each field with 13\farcm5 FoV
contains dozens of standard stars with high photometric precision.  There are six
fields observed by AST3-1 in 2012, which are at $\alpha=$ 0$^{\hbox{\scriptsize h}}$, 2$^{\hbox{\scriptsize h}}$, 20$^{\hbox{\scriptsize h}}$, 
22$^{\hbox{\scriptsize h}}$ and $\delta= -60\degr$, as well as the fields of JL~82 and NGC~458. The median 
magnitude differences between AST3-1 and the standard $i$ system in these 
fields are $-0.03$, $-0.06$,
$-0.01$, $-0.01$, 0.00 and $-0.04$, respectively.  This indicates that the absolute 
photometric calibration of AST3-1 has a RMS of about 0.02\,mag.
Besides, the $r - i$ colour coefficient in the $i$ magnitude transformation is 
around 0.2 mag.

\section{data access}

The AST3-1 DR1 dataset is available to the community through the Chinese 
Astronomical Data Center\footnote{\url{http://explore.china-vo.org/}}.  
The dataset consists of three parts: AST3-1 images, AST3-1 survey and AST3-1 light 
curves. 

AST3-1 images includes 14,460 corrected images and corresponding catalogues from 
both the survey and monitoring observations, accompanied 
with observation information such as center coordinates, date, exposure time 
and image quality (see Table~\ref{tab:db-image}). The AST3-1 survey is a combined 
catalogue containing more than 16 million sources (see Table~\ref{tab:survey}) 
down to $i \sim 19$
from the survey observations with positions and magnitudes (see 
Table~\ref{tab:db-survey}). The mean values of positions and magnitudes are adopted 
for the stars detected in multiple images (the median number of observations is three).  
AST3-1 light 
curves also contains the sources from the monitored fields, as well 
as nearly 2 million light curves for the sources with more than 50 detections (see 
Table~\ref{tab:monitoring}). Each light 
curve is presented as a PNG picture and available in multiple formats for download, 
including FITS binary table, CSV file and VOTable.
Table~\ref{tab:db-lc}) summaries the relative information stored in the FITS header.
In addition, we provide a tar file for each field containing all light curves in 
FITS, CSV and VOTable format, respectively (the link is at the top of the AST3-1 light 
curves webpage).

\begin{table*}
\begin{center}
\caption{AST3-1 image Database Schema}
\label{tab:db-image}
\begin{tabular}{ll}
\hline
\hline
Column Name  &  Description\\
\hline
Field & Field name\\
IMAGE & Image name and download link\\
CATALOGUE  & Catalogue  name and download link\\
DATE-OBS & UTC time of the beginning of observation\\
MJD & Modified Julian date of the beginning of observation\\
EXPTIME & Integration time (s)\\
TEMP\_CCD & Temperature of the CCD (K)\\
RA & Right Ascension J2000 of the image center (deg)\\
DEC & Declination J2000 of the image center (deg)\\
SKY & Background brightness (ADU)\\
FWHM & Median Full Width at Half Maximum across the FoV (pixel)\\
ELONGATION & Median ratio of semi-major to semi-minor axis across the FoV\\
\hline
\end{tabular}
\end{center}
\end{table*}

\begin{table*}
\begin{center}
\caption{AST3-1 survey Database Schema}
\label{tab:db-survey}
\begin{tabular}{ll}
\hline
\hline
Column Name  &  Description\\
\hline
ID & Source index\\
RA & Right Ascension J2000 (deg)\\
DEC & Declination J2000 (deg)\\
MAG & Mean magnitude (mag)\\
MAGERR & Standard deviation of magnitudes (mag)\\
COUNT & Number of observations\\
\hline
\end{tabular}
\end{center}
\end{table*}

\begin{table*}
\begin{center}
\caption{AST3-1 light curve Table Schema}
\label{tab:db-lc}
\begin{tabular}{ll}
\hline
\hline
Column Name  &  Description\\
\hline
DATE & UTC time of the beginning of observation\\
MJD & Modified Julian date of the beginning of observation\\
X & Windowed X position in CCD (pixel)\\
Y & Windowed Y position in CCD (pixel)\\
RA & Right Ascension J2000 (deg)\\
DEC & Declination J2000 (deg)\\
FLUX & Flux (ADU)\\
FLUX\_ERR & Flux error (ADU)\\
MAG\_AUTO & Magnitude in Kron aperture (mag)\\
MAGERR\_AUTO & Magnitude error in Kron aperture (mag)\\
BACKGROUND & Background brightness (ADU)\\
FWHM & Full Width at Half Maximum, assuming a Gaussian profile (pixel)\\
A & Semi-major axis length (pixel)\\
B & Semi-minor axis length (pixel)\\
THETA & Position angle of semi-major axis (degrees East from North)\\
FLAGS & SExtractor flags for the source\\
R50 & Radii enclosing half of total flux \\
MAG & 4\arcsec-aperture magnitude (mag)\\
MAGERR & 4\arcsec-aperture magnitude error (mag)\\
\hline
\end{tabular}
\end{center}
\end{table*}

\section{Summary}

In 2012, the first AST3 telescope, AST3-1, was deployed at Dome A in Antarctica to 
carry out time-domain surveys. During the commissioning phase, AST3-1
surveyed $\sim$ 2000 deg$^2$ fields as well as the LMC and SMC, and monitored a dozen  
fields including the LMC center, a field for studying exoplanet transits and some Wolf-Rayet stars.  
After the raw data were 
returned to China, we performed aperture photometry, calibrated the magnitudes 
and produced light curves in the monitored fields.  In this paper we present the first data release, DR1, of 
the photometric data from the AST3-1 commissioning surveys.  DR1 consists of 14 thousand
scientific images, 16 million sources bright than $i \sim 19$ with photometry and 
astrometry, and 2 million light curves.

For faint sources, the $5\sigma$ limiting magnitude is $i \sim 18.7$ in 60\,s with a
typical FWHM of 3\farcs7 and dramatically deepens to $i \sim 19.3$ with a sharper FWHM of
2\farcs7.  As a result, we infer a depth of $i \sim 20$ in 60\,s during a
moonless night with FWHM of 2\arcsec, which is the approximate median ground layer seeing.
Along with its wide FoV, AST3-1 is capable of quickly discovering faint transients
such as Near-Earth Asteroids, SNe and optical counterparts of GRBs and
gravitational wave events. For example, the second AST3 participated in the follow-up of 
the first electromagnetic counterpart to a gravitational wave signal GW170817 and 
detected its quickly fading \citep{Hu2017}.  At the bright end, ($i \approx 11$), 
AST3-1 achieved photon noise limited precision of 0.8\,mmag, measured from two
consecutive 60\,s exposures. The AST3-1 photometry in light curves spanning one to two months is stable 
to a level of 5\,mmag, mainly limited by variations in observational conditions and system
stability.  More careful de-trending and binning techniques can further reduce the 
systematic and statistical uncertainties.  Therefore the dataset can be very useful in stellar
variability research and the data in some fields are possible to be used for 
exoplanet detection.  Moreover, our results
provide suggestions for future strategies, such as choosing the integration time and FWHM
for a given magnitude to optimize the precision.  

In conclusion, the commissioning of AST3-1 has confirmed its promising prospects 
in time-domain astronomy, taking advantage of the clear and dark polar night at 
Dome A. The dataset from commissioning is public to the community through the 
Chinese Astronomical Data Center.

\section*{Acknowledgments}

The authors deeply appreciate the 28th and 29th CHINARE for their great 
effort in installing/maintaining AST3-1 and PLATO-A.
This study has been supported by the National Basic Research Program (973
Program) of China (Grant No.\ 2013CB834900) and the Chinese Polar
Environment Comprehensive Investigation $\&$ Assessment Program (Grant No.\
CHINARE2016-02-03), the National Natural Science Foundation of China 
(NSFC) (Grant Nos.\  11403057, 11403048, 11203039 and 11273019). 
DWF and BLH acknowledge the support from NSFC (Grant No.\ 11503051), the Joint 
Research Fund in Astronomy (U1531115, U1731243) under cooperative agreement 
between the NSFC and Chinese Academy of Sciences (CAS), the National R\&D 
Infrastructure and Facility Development Program of China, ``Earth System 
Science Data Sharing Platform" and ``Fundamental Science Data Sharing Platform" 
(DKA2017-12-02-XX). Data resources are supported by Chinese Astronomical Data 
Center (CAsDC) and Chinese Virtual Observatory (China-VO). 
S.W. thanks the Heising-Simons Fundation for their generous support. 
Y.Y. acknowledge the support from a Benoziyo Prize Postdoctoral Fellowship.
ZHZ acknowledge the support from the 973 Program (Grant No.\ 2014CB845800), 
the NSFC (Grants Nos.\ 11633001 and 11373014), the Strategic Priority
Research Program of the CAS (Grant No. XDB23000000) and 
the Interdiscipline Research Funds of Beijing Normal University.

The construction of the AST3 telescopes was made possible by funds from
Tsinghua University, Nanjing University, Beijing Normal University,
University of New South Wales, Texas A$\&$M University, the Australian
Antarctic Division and the National Collaborative Research Infrastructure
Strategy (NCRIS) of Australia. It has also received funding from the
Chinese Academy of Sciences through the Center for Astronomical
Mega-Science and National Astronomical Observatories (NAOC).
This research was made possible through the use of the AAVSO Photometric All-Sky
Survey (APASS), funded by the Robert Martin Ayers Sciences Fund.

\bsp	
\label{lastpage}
\end{document}